\documentstyle[12pt]{article}
\begin{document}
\vspace*{-1in}
\renewcommand{\thefootnote}{\fnsymbol{footnote}}
\begin{flushright}
TIFR/TH/96-50\\
hep-ph/9609285\\
\end{flushright}
\vskip 65pt
\begin{center}
{\Large \bf ${}^1P_1$ charmonium production at the Tevatron} \\
\vspace{8mm}
{\bf K. Sridhar\footnote{sridhar@theory.tifr.res.in}}\\
\vspace{10pt}
{\it Theory Group, Tata Institute of Fundamental Research, \\ 
Homi Bhabha Road, Bombay 400 005, India.}

\vspace{80pt}
{\bf ABSTRACT}
\end{center}
The production of the ${}^1P_1$ charmonium state, $h_c$, at large-$p_T$
at the Tevatron is considered. The colour-octet contributions to
this state are found to be dominant and give a reasonably large
rate for the production of $h_c$. This should make it feasible to 
look for this resonance in the $J/\psi +\pi$ decay channel.
This production rate is a prediction of NRQCD, and the observation of
the $h_c$ at the Tevatron can, therefore, be used as a test of the 
colour-octet predictions of NRQCD.
\vspace{12pt}

\vspace{98pt}
\noindent
\begin{flushleft}
September 1996\\
\end{flushleft}

\vskip 10pt

\setcounter{footnote}{0}
\renewcommand{\thefootnote}{\arabic{footnote}}

\vfill
\clearpage
\setcounter{page}{1}
\pagestyle{plain}
The production of quarkonia has conventionally been described in
terms of the colour-singlet model \cite{berjon, br}, which uses
a non-relativistic approximation to describe the binding
of the heavy-quark pair produced via parton-fusion processes into
a quarkonium state. The heavy-quark pair is projected onto a 
physical quarkonium state using a colour-singlet projection and 
an appropriate spin-projection. The failure of this model in 
explaining the data on the inclusive large-$p_T$ $J/\psi$ production 
cross-section measured by the CDF experiment \cite{cdf} has led to 
a major revision of the theory of quarkonium formation. The colour-singlet 
model is an approximation which works well only when the relative velocity
$v$ between the quarks in the quarkonium state can be completely neglected.
It is possible to include the effects coming from higher orders in a 
perturbation series in $v$ in the framework of an effective
field theory called non-relativistic QCD (NRQCD) \cite{caswell}. 

NRQCD is an effective field theory derived from the full QCD
Lagrangian by neglecting all states of momenta larger than
a cutoff of the order of the heavy quark mass, $M$, \cite{bbl} and 
accounting for this exclusion by introducing new interactions in the 
effective Lagrangian, which are local since the excluded states
are relativistic. Beyond the leading order in $1/m$ the effective theory is
non-renormalisable. The physical quarkonium states can then be expanded 
in terms of its Fock-components in a perturbation series in $v$, and
it then turns out that the $Q \bar Q$ states appear in either
colour-singlet or colour-octet configurations in this series.
Of course, the physical state is a colour-singlet so that a 
colour-octet $Q \bar Q$ state is connected to the physical
state by the emission of one or more soft gluons. In spite of
the non-perturbative nature of the soft gluons emitted, it
turns out that the effective theory still gives us some
useful information about the intermediate octet states. This is 
because the dominant transitions from colour-octet to physical
colour-singlet states are $via$ $E1$ or $M1$ transitions with
higher multipoles being suppressed by powers of $v$. It then becomes
possible to use the usual selection rules for these radiative transitions
to keep account of the quantum numbers of the octet states, so that
the short-distance coefficient corresponding to the octet state
can be calculated and its transition to a physical singlet state
can be specified by a non-perturbative matrix element. The cross-section
for the production of a meson $H$ then takes on the following factorised form:
\begin{equation}
   \sigma(H)\;=\;{\rm Im}\,\sum_{n=\{\alpha,S,L,J\}} {F_n\over m^{d_n-4}}
       \langle{\cal O}^H_\alpha({}^{2S+1}L_J)\rangle
\label{e1}
\end{equation}
where $F_n$'s are the short-distance coefficients and ${\cal O}_n$ are local
4-fermion operators, of naive dimension $d_n$, describing the long-distance
physics. The cutoff-dependence of $F_n$ is compensated by that of the 
long-distance matrix elements. 

The importance of colour-octet components was seen \cite{jpsi} in the
phenomenology of $P$-state charmonium production at large $p_T$ in 
the Tevatron data. These processes do not have a consistent 
description in terms of colour singlet operators alone \cite{bbl2}. 
However, even for the direct production of $S$-states
such as the $J/\psi$ or $\psi'$, where the colour singlet components give
the leading contribution in $v$, the inclusion of sub-leading octet states
was seen to be necessary for phenomenological reasons \cite{brfl}. 
The long-distance matrix elements are not calculable, and 
a linear combination of octet matrix-elements have been fixed by 
fitting to the Tevatron data \cite{cho}.
Independent tests of the $S$-state colour octet enhancement are
important and recent work shows that a different linear combination
of the same colour octet-matrix elements that appear in the Tevatron
analysis also appears in the analyses of photoproduction \cite{photo}
and hadroproduction experiments \cite{hadro}. These analyses provide
an important cross-check on the colour-octet contributions.
Other production modes like $J/\psi$ production in low-energy
$e^+ - e^-$ collisions \cite{brch}, on the $Z$-peak at
LEP \cite{lep} and in $\Upsilon$ \cite{upsilon} decays have been 
considered, with the purpose of determining the magnitude of the
octet contribution.

In this letter, we compute the large-$p_T$ production cross-section
for the ${}^1P_1$ charmonium state, $h_c$. The production of this 
resonance is interesting in its own right~: charmonium spectroscopy 
\cite{onep} predicts this state to exist at the centre-of-gravity 
of the $\chi_c ({}^3P_J)$ states. The E760 collaboration at the 
Fermilab have reported \cite{e760} the first observation of this
resonance but this needs further confirmation. It is, therefore, 
interesting to determine the theoretical production rate for this
resonance at the Tevatron and to know whether there is any chance
that its existence can be confirmed. From our point of view the 
production of the ${}^1P_1$ is interesting for yet another reason. 
Being a $P$-state, the leading colour-singlet contribution is
already at $O(v^2)$, and at the same order we have the octet
production of the ${}^1P_1$ state through an intermediate 
${}^1S_0$ state. As we shall see later, we can infer the 
non-perturbative matrix elements for the production of $h_c$
from the matrix-elements of other states. One channel which may 
be suitable for the detection of the $h_c$ is its decay into a 
$J/\psi +\pi$. Armed with the non-perturbative matrix elements we can 
$predict$ the number of $J/\psi +\pi$ events in the total 
sample of $J/\psi$'s accumulated by the Tevatron experiments, 
provided we know the decay branching fraction for $h_c \rightarrow 
J/\psi + \pi$. This latter number has been estimated from spectroscopy
We believe that with these inputs, a reasonably 
solid prediction of the production cross-section of the 
${}^1P_1$ charmonium state is possible within the NRQCD framework. 
It is worth mentioning again that we
are considering the large-$p_T$ production of this state and
therefore, we are on firm grounds in using the factorisation
relation of the kind shown in Eq.~\ref{e1}.

The subprocesses that we are interested in are the following:
\begin{eqnarray}
g + g \rightarrow {}^1P_1^{\lbrack 1 \rbrack} + g, \nonumber \\
g + g \rightarrow {}^1S_0^{\lbrack 8 \rbrack} + g, \nonumber \\
q(\bar q)+ g \rightarrow {}^1S_0^{\lbrack 8 \rbrack } + 
                  q(\bar q), \nonumber \\
q + \bar q \rightarrow {}^1S_0^{\lbrack 8 \rbrack } + g.
      \label{e2}
\end{eqnarray}
The ${}^1S_0^{\lbrack 8 \rbrack} \rightarrow h_c$ is mediated by
a gluon emission in a $E1$ transition. The large-$p_T$ hadronic 
production cross-section is given as
\begin{eqnarray}
&&{d\sigma \over dp_T}(p \bar p \rightarrow {}^1P_1 X)
=  \nonumber \\
&& \sum \int dy\int dx_1 x_1G_{a/p}(x_1) x_2G_{b/\bar p}(x_2)
 {4p_T \over 2x_1 -\bar x_T e^y}
 {d\hat \sigma \over d \hat t}(ab \rightarrow {}^2S+1L_J c) .
\label{e3}
\end{eqnarray}
In the above expression, the sum runs over all the initial partons
contributing to the subprocesses;
$G_{a/p}$ and $G_{b/\bar p}$ are the distributions of partons 
$a$ and $b$ in the hadrons with momentum fractions $x_1$ and
$x_2$, respectively. Energy-momentum conservation determines
$x_2$ to be
\begin{equation}
x_2= {x_1 \bar x_T e^{-y} - 2 \tau \over 2x_1-\bar x_T e^y},
\label{e4}
\end{equation}
where $\tau = M^2/s$, with $M$ the mass of the resonance, $s$
the centre-of-mass energy and $y$ the rapidity at which the resonance
is produced.
\begin{equation}
\bar x_T= \sqrt{x_T^2 + 4\tau} \equiv {2M_T \over \sqrt{s}},
\hskip20pt x_T={2p_T \over \sqrt{s}}
\label{e5}
\end{equation}
The expressions for the singlet and the octet subprocess cross-sections,
$d\hat\sigma/d\hat t$, are given in Refs.~\cite{gtw} and \cite{cho},
respectively. 

The production cross-section for the ${}^1P_1$ state is fully
specified, once we have specified the colour-singlet matrix
element for the ${}^1P_1$ state 
$\left\langle{\cal O}^{h_c}_1({}^1P_1)\right\rangle$
and the value for the colour-octet
matrix element that takes the octet ${}^1S_0$ state to a $h_c$,
$\left\langle{\cal O}^{h_c}_8({}^1S_0)\right\rangle$.
We note that the matrix element of the singlet ${}^1P_1$ state is 
related to the derivative of the wavefunction of at the origin by
\begin{equation}
   \left\langle{\cal O}^{h_c}_1({}^1P_1)\right\rangle
      \;=\;{27\over2\pi}|R'(0)|^2.
\label{e6}
\end{equation}
The Tevatron data on $\chi_c$ production fixes \cite{cho} the 
colour-octet matrix element which specifies the transition of a 
${}^3S_1$ octet state into a ${}^3P_J$ state. We would expect from 
heavy-quark spin symmetry of the NRQCD Lagrangian \cite{bbl} that the 
matrix-element for ${}^1S_0^{\lbrack 8 \rbrack} \rightarrow h_c$ should be
of the same order as that for ${}^3S_1^{\lbrack 8 \rbrack} \rightarrow 
{}^3P_1$. This is because the essential difference between these
transitions comes through the magnetic quantum number so that
the corrections to this equality will be of $O(v^2) \sim$
30\%. For the derivative of the wave-function we use a similar
argument to fix it to be the same as for the $\chi_c$ states. 
 
We have computed the cross-sections for the Tevatron energy
$\sqrt{s}=1.8$~TeV. In Fig.~1, we present the ${}^1P_1$ production 
cross-section $d\sigma / dp_T$ as a function of $p_T$, where the  
cross-section has been folded in with the branching ratio of the
${}^1P_1$ state into $J/\psi+\pi$. We have integrated over the  
full rapidity interval $-0.6 \le y \le 0.6$ covered by the CDF 
experiment. For the results shown in Fig.~1, we have used the
MRSD$-'$ densities \cite{mrs}. The parton densities are evolved to a 
scale $Q = M_T/2$. For the singlet matrix element, we use the
value extracted from $\chi_c$ decays, which is 
$\left\langle{\cal O}^{h_c}_1({}^1P_1)\right\rangle=0.32$ 
\cite{mangano} and for the octet matrix element we have
$\left\langle{\cal O}^{{}^1P1}_8({}^1S_0)\right\rangle=0.0098$ 
\cite{cho}. With these inputs, we find that the cross-section for 
$h_c$ production (folded in with the decay fraction into a 
$J/\psi$ and $\pi$, which we take to be 0.5\% \cite{onep}) 
integrated over the region between 5 and 20~GeV 
in $p_T$ is quite substantial. With the 20~pb${}^{-1}$ total luminosity
accumulated at the Tevatron, we expect of the order of 650
events in the $J/\psi + \pi$ channel. Of this the
contribution from the colour-singlet channel is a little
more than 40, while the octet channel gives more than 600 events.
The colour-octet dominance is more pronounced at 
large-$p_T$. The octet contribution has a flatter $p_T$ dependence
as compared to the singlet which falls off rather rapidly with
$p_T$. The shape of the $p_T$ distribution is also, therefore, a
testable prediction of NRQCD.

We have studied the effect on the cross-section of the variation 
of the parton densities, the scale and the non-perturbative 
matrix-elements. By using GRV densities \cite{grv} instead of
MRSD$-'$, we find that the cross-section increases by about 20-25\%.
Varying the scale from $M_T/2$ to $2M_T$ reduces the predictions
by roughly a factor of 2. A 25\% variation results if we vary the
colour-octet matrix element by 30\% around the central-value quoted above.
The decay branching fraction of $h_c$ into a $J/\psi+\pi$ could
be as large as 1\%, and if we use this instead of the 0.5\% used
in the above calcuations we could have a production cross-section
which is twice as large.

Before we conclude, a few remarks about this remarkable prediction
of NRQCD is in order. There is now sufficient empirical evidence
for the failure of the colour-singlet model and the need to include
effects beyond this model. NRQCD provides a suitable way of going
beyond the colour-singlet model and, as discussed, has had considerable
success in describing a large amount of data on quarkonium decay and
production. There however exists a different approach to quarkonium
production the so-called colour-evaporation model or, alternatively
called, the semi-local duality model \cite{semi}. The idea in this
model is to relate the integral of the open-charm production 
cross-section to the sum of the resonance production cross-sections.
The individual resonance cross-sections are not predicted but are
obtained by fitting a parameter to experimental data. Detailed results
of this model for hadroproduction of $J/\psi$ have been obtained
recently \cite{gavai, halzen} and the comparison with the data
is as successful as that of NRQCD. But we must bear in mind that
NRQCD is a more predictable model. In particular, when some of the
non-perturbative parameters have been determined from experiment
it is possible to use symmetry relations to estimate other 
non-perturbative parameters. This is true of the production 
process under consideration in this paper. We are, therefore,
able to predict the number of $h_c$ events that we will see at
the Tevatron, which is not predicted in any semi-local duality
based approach, where the undetermined parameters can only be
fitted $a\ posteriori$.

In conclusion, we find that a reasonably large rate for the production
of the ${}^1P_1$ is expected at the Tevatron, with a dominant contribution
from the colour-octet production channel. Heavy-quark symmetry relations
allow us to infer the size of the non-perturbative matrix-elements and 
to predict the rate for $h_c$ production and its subsequent
decay into a $J/\psi$ and a $\pi$. By looking at $J/\psi$ events associated
with a soft pion, it should be possible to pin down the elusive ${}^1P_1$
resonance at the Tevatron, given the large number of $J/\psi$ events
already available. We believe that this is a firm prediction of the
NRQCD framework and may be a useful way of distinguishing this from
other models of quarkonium formation.

\newpage
\begin{figure}[htb]
\vskip 8in\relax\noindent\hskip -1in\relax{\includegraphics{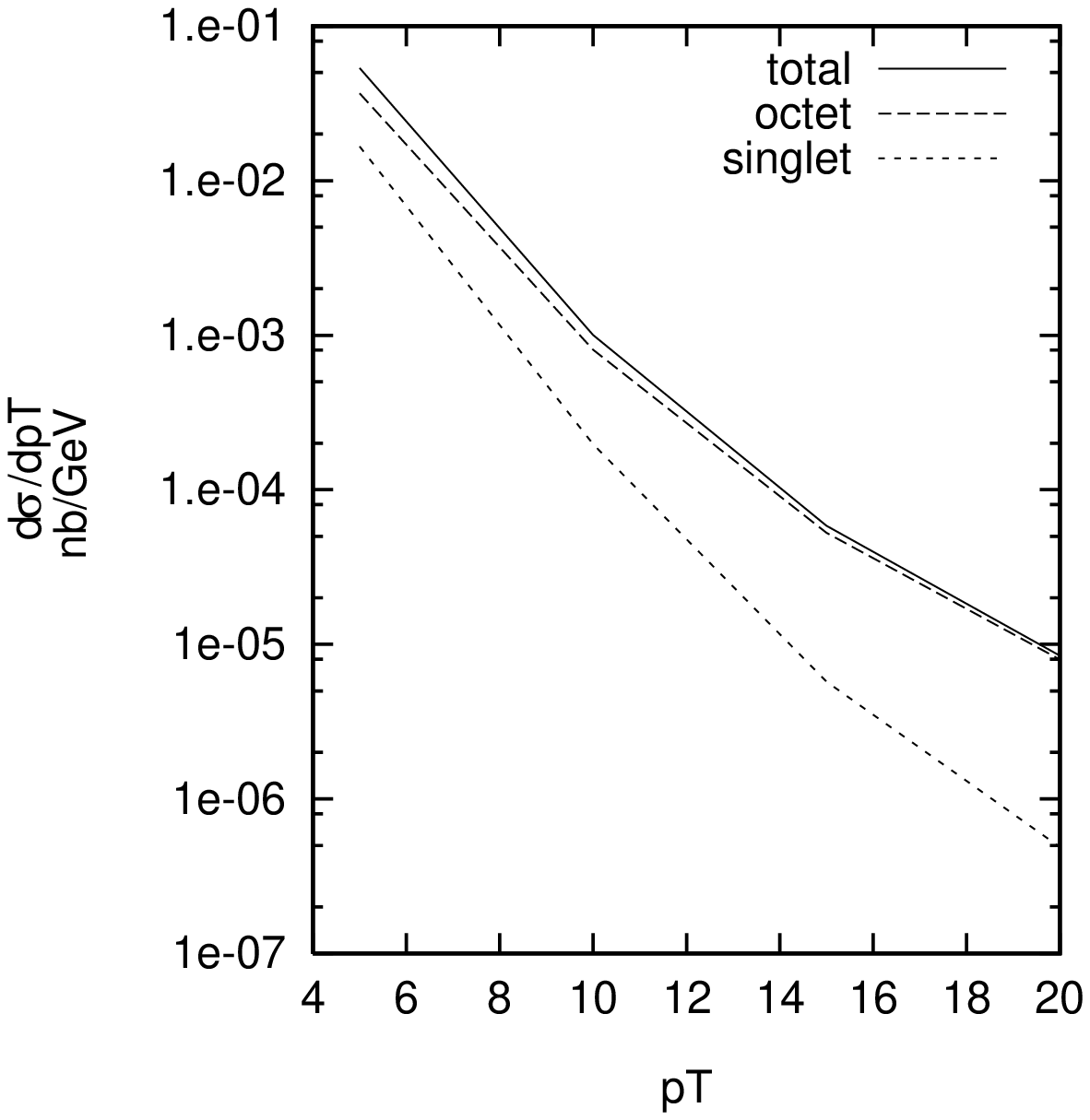}}

\vspace{-20ex}
\caption{$d\sigma/dp_T$ (in nb/GeV) for $h_c$ production (after folding in
the decay branching fraction of the $h_c$ into a $J/\psi+\pi$) at
1.8~TeV c.m. energy with $-0.6 \le y \le 0.6$. The colour-singlet, 
colour-octet and the total contributions are shown.}
\end{figure}
\end{document}